**Impact-induced melting during accretion of the Earth**

Jellie  de Vries[1]

Email: Jellie.bgi@gmail.com

Francis Nimmo[2]

Email: fnimmo@es.ucsc.edu

H. Jay Melosh[3]

Email: jmelosh@purdue.edu

Seth A. Jacobson[1,4]

Email: seth.jacobson@oca.eu

Alessandro Morbidelli[4]

Email: morby@oca.eu

David C. Rubie[1]

Corresponding author

Email: dave.rubie@uni-bayreuth.de

(Institutional addresses)

[1] Bayerisches Geoinstitut, University of Bayreuth, Universitätsstraße 30, 95440, Bayreuth, Germany

[2] Department of Earth and Planetary Sciences, University of California Santa Cruz, 1156 High Street,

Santa Cruz, CA 95064, USA

[3] Department of Earth, Atmospheric and Planetary Sciences, Purdue University, 550 Stadium Mall Drive, West Lafayette, IN 47907, USA

[4] Observatoire de la Côte d'Azur, Boulevard de l'Observatoire, 06304 Nice, France

## Abstract

Because of the high energies involved, giant impacts that occur during planetary accretion cause large degrees of melting. The depth of melting in the target body after each collision determines the pressure and temperature conditions of metal-silicate equilibration and thus geochemical fractionation that results from core-mantle differentiation. The accretional collisions involved in forming the terrestrial planets of the inner Solar System have been calculated by previous studies using N-body accretion simulations. Here we use the output from such simulations to determine the volumes of melt produced and thus the pressure and temperature conditions of metal-silicate equilibration, after each impact, as Earth-like planets accrete. For these calculations a parametrised melting model is used that takes impact velocity, impact angle and the respective masses of the impacting bodies into account. The evolution of metal-silicate equilibration pressures (as defined by evolving magma ocean depths) during Earth's accretion depends strongly on the lifetime of impact-generated magma oceans compared to the time interval between large impacts. In addition, such results depend on starting parameters in the N-body simulations, such as the number and initial mass of embryos. Thus, there is the potential for combining the results, such as those presented here, with multistage core formation models to better constrain the accretional history of the Earth.

## Background

Accretion of the Earth took place over a time period on the order of 100 million years (e.g. Jacobson et al., 2014). According to current astrophysical theories, the Earth and the other terrestrial planets formed in three stages: (1) In the protoplanetary disk, dust condensed from a cooling gas of solar composition. Dust grains settled to the midplane and coagulated to form approximately 100-km-size solid "planetesimals" likely through gravoturbulent instabilities (see Johansen et al. 2015 for a review). (2) From this sea of planetesimals, planetary embryos (lunar- to Mars-mass bodies) emerged from runaway growth processes either due to gravitational focusing (Greenberg et al., 1978) or pebble accretion (Lambrechts and Johansen, 2012). The runaway growth phase was then followed by one of oligarchic growth (Kokubo and Ida, 1998) while the system of embryos stabilized dynamically via dynamical friction (see Jacobson & Walsh 2015 for a review). (3) The final stage of accretion was dominated by the mutual gravitational interactions of the embryos and was characterized by large, violent collisions (Benz et al., 1989). This final stage has been the subject of many recent modelling studies (e.g. Chambers and Weatherill, 1998; Chambers, 2001, 2013; O'Brien et al., 2006; Walsh et al. 2011; Jacobson and Morbidelli, 2014).

A major differentiation process that occurred during the early history of the Earth led to the formation of its iron-rich core and silicate mantle. Core formation involved the separation of metal from silicates for which high temperatures were essential. Most likely both the metal and the silicate had to be in a molten state for core formation to occur efficiently (Stevenson 1990; Rubie et al., 2003, 2015a). Following the decay of short-lived $^{26}$Al, the heat required for core-mantle differentiation of the Earth resulted primarily from high-energy impacts with other planetary bodies. For example, the postulated impact between the proto-Earth and a large (e.g. possibly Mars-size) impactor traveling at a velocity on the order of 10 km/s, which is considered to have been responsible for the formation of the Moon (e.g. Hartmann and Davis, 1975; Canup and Asphaug, 2001; Ćuk and Stewart, 2012; Canup, 2012), may have provided sufficient energy to melt the entire Earth (Fig. 3 in Rubie et al., 2015a; Nakajima and Stevenson, 2015).

Geochemical models of core formation can be used to constrain the early melting history of the Earth. The segregation of liquid metal from silicate during formation of the Earth's core depleted siderophile (metal-loving) elements in the Earth's mantle by transporting them into the core. Thus elements such as Fe, Ni, Co, Mo, W, V and Cr are depleted in the mantle by up to 2 orders of magnitude relative to their abundances in CI chondrites (e.g. Rubie et al., 2015a). The degree of the depletion of moderately siderophile elements is variable and depends on the metal-silicate partition coefficient which, for element M, is defined as the ratio of the molar concentration of M in the metal to its concentration in the silicate:

$$D_{\mathrm{M}}^{met-sil} = C_{\mathrm{M}}^{met} / C_{\mathrm{M}}^{sil}.$$

The partition coefficient $D_{\mathrm{M}}^{met-sil}$ depends on pressure ($P$), temperature ($T$), oxygen fugacity ($f_{\mathrm{O2}}$) and, for some elements, the compositions of the metal and silicate phases. Because $D_{\mathrm{M}}^{met-sil}$ depends on $P$ and $T$, the conditions of core formation have been estimated by matching experimentally-determined metal-silicate partition coefficients with core-mantle partition coefficients determined from McDonough (2003). For example, Li and Agee (1996) showed that to match $D_{\mathrm{M}}^{met-sil}$ for Ni and Co with core-mantle partition coefficients requires an equilibration pressure of ~28 GPa. Subsequent studies have derived conditions of core formation that range from 27 to 60 GPa and 2200 to 4200 K (see Table 3 in Rubie et al., 2015a). However, this approach is based on the concept of "single-stage core formation" whereby all the metal of the core is considered to equilibrate with all silicate of the mantle at some mid-mantle depth (e.g. Corgne et al., 2009; Righter 2011; Walter and Cottrell, 2013). In reality, the Earth accreted through a series of high-energy impacts with smaller bodies consisting of multi-km-size planetesimals (possibly hundreds of km in diameter) and Moon- to Mars-size embryos (e.g. Chambers and Wetherill, 1998). Most impacts, while delivering the energy that caused melting, added Fe-rich metal that segregated to the Earth's proto-core. Thus core formation was multistage and occurred as an integral part of the accretion process (Wade and Wood, 2005; Rubie et al., 2011).

The preliminary multistage core formation model of Rubie et al. (2011) has been integrated recently with N-body accretion simulations (Rubie et al., 2015b). Hundreds of impacts and associated core-formation events were simulated for all embryos as well as for the final terrestrial planets. A mass balance/element partitioning approach was used, which required the bulk compositions of all accreting bodies to be defined. Elements were assumed to be present mostly in Solar System (CI) relative abundances and the oxygen content was the main compositional variable. Metal-silicate equilibration pressures and bulk compositional variables were refined by least squares minimization in order to fit the calculated composition of Earth's mantle to that published for the primitive mantle (Palme and O'Neill, 2013); equilibration temperature was set midway between the peridotite solidus and liquidus at the equilibration pressure. A simplifying assumption is that metal-silicate equilibration pressures were a constant fraction of the core-mantle boundary pressure at the time of each impact, irrespective of the mass of the impactors. As constrained by the least squares fits to Earth's mantle composition, this fraction is 0.6-0.7. Thus, pressures increased as the Earth's mass increased during accretion and reached a maximum of ~80 GPa.

It has been assumed, in almost all studies of siderophile element partitioning, that the pressure of metal-silicate equilibration during core formation corresponds to the pressure at the base of the magma ocean, with temperature being defined by the equivalent peridotite liquidus or solidus. In the case of dispersed metal droplets that continuously re-equilibrated as they sink in a magma ocean, more realistic fractionation models have been proposed according to which the effective equilibration pressure differs significantly from the pressure at the base of a magma ocean (Rubie et al., 2003). However, accretion of Earth mostly involved collisions with bodies that previously had undergone core-mantle differentiation (e.g. Kleine et al., 2009). In this case, it is likely that the core of the impactor would sink through the magma ocean entraining silicate liquid to form a descending, expanding high-density plume of mixed metal and silicate (Deguen et al., 2011; Rubie et al., 2015b). The hydrodynamic model formulated by Deguen et al. (2011) to describe this plume has two important consequences for modelling core formation: (1) Only a small fraction of the silicate mantle of the target body equilibrates with the accreted metal during each impact (Rubie et al., 2015b). (2) The pressure of final metal-silicate equilibration corresponds to the pressure near the base of the magma ocean where the descending metal-silicate plume comes to rest. Based on this latter point, it is clear that understanding the evolution of magma ocean depth during accretion is of fundamental importance for modelling core-mantle differentiation.

Complete chemical equilibrium between metal and silicate during core formation requires that the iron emulsifies into small (cm size) droplets (Rubie et al., 2003). In the case of small impacting bodies (planetesimals), emulsification is likely to be complete but for embryos the extent of emulsification is much less certain (Dahl and Stevenson, 2010; Samuel, 2012; Deguen et al., 2014; Kendall and Melosh, 2016). The extent of emulsification and the fraction of metal that equilibrates is of major importance for understanding the origin of planetary tungsten isotopic anomalies (Nimmo et al., 2010; Rudge et al., 2010) and also affects the final distribution of siderophile elements (Rubie et al., 2015b).

**Aims of the study**

The aim of this contribution is to present the results of preliminary calculations of melting depths caused by each individual impact in N-body accretion simulations and to develop a complete history of melting and evolving magma ocean depth throughout Earth's accretion history. Because magma ocean depths are related to the conditions of metal-silicate equilibration, this approach could eventually be used to input equilibration pressures and temperatures into the multistage accretion/core formation model of Rubie et al. (2015). This would enable the assumption that equilibration pressures are a constant fraction of core-mantle boundary pressures to be superseded. The approach may also enable the most realistic accretion models to be identified.

**Model description**

The volume of melt produced during an accretional collision depends on the kinetic energy of the impactor, as determined by its mass and velocity, and the impact angle which determines how much of the kinetic energy is transferred into the target body. These three parameters (impactor mass, impact velocity and impact angle) and the mass of the target body are determined for each impact in the N-body accretion simulations. However, since the simulations involve 1000 to 2000 impacts, it would be much too time consuming to perform full three-dimensional hydrocode calculations of the melt volume produced by each impact (Marinova et al. 2011). Two-dimensional models also cannot be used for non-vertical impacts due to their assumed symmetry in the third dimension. We have therefore based our calculations on the analytical model of Bjorkman and Holsapple (1987), who constructed a simple relationship between melt volume, impactor mass and impact velocity for vertical impacts using dimensional analysis. Based on scaling theory and hydrocode simulations (Pierazzo and Melosh, 2000), a dependence on the impact angle has been included to cover the whole range of impact parameters that occur during planetary accretion (Abramov et al., 2012). Although our approach is undoubtedly simplified, it reveals general trends (such as the importance of magma ocean freezing time) that are unlikely to change significantly if more sophisticated approaches were to be implemented. The simple approach adopted here also makes the dependence of the results on input parameters – which are often poorly-known – more transparent.

We have used this model to calculate the melt volume produced by 1000-2000 impacts that occur during each of three "Grand Tack" N-body accretion simulations (Walsh et al., 2011; Jacobson & Morbidelli, 2014) and have determined the corresponding metal-silicate equilibration pressures at the base of the resulting melt pools and global magma oceans.

The chosen N-body simulations are generalizable examples because both classical and Grand Tack terrestrial planet formation scenarios grow Earth from a series of planetesimal and embryo impacts. We chose to consider only simulations from the Grand Tack model because the latter retains the successes of prior models while also creating a low mass Mars, explaining the compositional and dynamical structure of the asteroid belt (Walsh et al., 2011), and delivering water to the Earth (O'Brien et al., 2014; Rubie et al., 2015b). The simulations are from Jacobson & Morbidelli (2014) and Rubie et al. (2015) and are chosen because they reproduce well the mass and orbital characteristics of Earth (see also below).

## Melt volume

According to the analytical model of Bjorkman and Holsapple (1987), the mass of melt $M_{melt}$ produced by a vertical impact is given by:

$$\frac{M_{melt}}{m_p} = k \left( \frac{v^2}{E_m} \right)^{3\mu/2}. \qquad (1)$$

Here $m_p$ is the mass of the projectile, $v$ is the impact velocity, $k$ and $\mu$ are scaling parameters and $E_m$ is the energy required to melt the target upon decompression (see also Table 1). The values of constants in this relationship have been constrained through laboratory experiments (e.g. Schmidt and Housen, 1987) and through numerical simulations (e.g. Elbeshausen et al., 2009; Barr and Citron, 2011).

By including a dependence on impact angle (Pierazzo and Melosh, 2000), the analytical model has been modified by Abramov et al. (2012) to give:

$$V_{melt} = \frac{\pi}{6} k E_m^{-3\mu/2} \frac{\rho_p}{\rho_t} D_p^3 v^{3\mu} \sin^{2\gamma} \theta \qquad (2)$$

Here $V_{melt}$ is the melt volume, $\rho_p$ and $\rho_t$ are the densities of the projectile and target respectively, $D_p$ is the projectile diameter, $\gamma$ is a scaling parameter and $\theta$ is the impact angle (which for a vertical impact is defined as 90°). For simplicity, we assume that the mean densities of the bodies scale linearly with their mass, which is consistent with an increased self-compression for a larger body (Rubie et al., 2011).

If the target is already at elevated temperature (e.g. due to impacts), the energy required for melting is reduced. To take this into account, $E_m$ is calculated when the target is already at high temperature from:

$$E_m = E_m^0 \left( 1 - \frac{C_p \left( T_s + \frac{dT}{dz} d_m \right)}{C_p \left( T_{l0} + \frac{dT_l}{dP} P \right) + L_m} \right) \qquad (3)$$

(adapted from Abramov et al., 2012), where $T_s$ is the surface temperature prior to an impact and $T_{l0}$ is the liquidus temperature at the surface (1950 K); the temperature, $T$, and liquidus temperature, $T_l$, are calculated at the maximum depth of melting, $d_m$. For the liquidus temperature this is done through a temperature increase as a function of pressure, $P$. $C_p$ denotes specific heat and $L_m$ is the latent heat of melting. In the calculations described below, the surface temperature is set to 1750 K (Table 1) for a solid body before impact and the temperature increase with depth (0.1 K/km, based on a lower-mantle temperature gradient – Monnereau and Yuen, 2002), is assumed to be linear. Values of the material properties used here are listed in Table 2.

The depth of an impact-induced melt pool is calculated by solving Eqs. 2 and 3 numerically based on assuming that the melt pool has a specific geometry. Here we assume a spherical geometry (see Fig. 1a) which, in contrast to a hemispherical geometry, takes account of the free surface and closely approximates actual melt pool geometries (Pierazzo et al., 1997). The pressure and temperature at the base of the melt pool can then potentially be used as the conditions of metal-silicate equilibration during an episode of core formation.

Each body in the N-body simulations is assumed to be differentiated into an iron core and a dunite mantle, where the core size is calculated from the mean planetary density, assuming a metal mass fraction of 0.34 and a core density that is a factor 2.5 larger than the mantle density. The core temperature of the target body depends on the deep mantle temperature with a temperature jump ΔT at the core-mantle boundary (CMB) of 100 K. Equations are first solved numerically using mantle parameters. If the depth of melting extends into the core, the equations are solved again in the core domain, with core values for all parameters, to determine the extent of core melting. Due to the spherical geometry of the melt pool, the amount of silicate mantle melting increases when the depth of melting extends deeper into the core.

Magma ocean evolution

If sufficient melt is produced as a spherical melt pool by an accretional impact, isostatic readjustment can cause the melt to spread over the surface of the target body to form a global magma ocean as depicted in Fig. 1 (Tonks and Melosh, 1992). This requires extensive solid-state deformation of the crystalline mantle material that is located below and adjacent to the melt pool (Fig. 1a). Reese and Solomatov (2006) have shown that the formation timescale of a global magma ocean is controlled by radial relaxation of solid mantle, because this timescale is much longer than the timescale of lateral melt flow. The radial relaxation timescale for a Mars-size body has been estimated by Reese and Solomatov (2006) and varies by 5-6 orders of magnitude depending on the volume of the impact-induced melt pool and the rheology of the underlying crystalline mantle. For small amounts of melting, the timescales are on the order of $10^4$-$10^5$ years (Reese and Solomatov, 2006). It is therefore unlikely that planetesimal impacts result in a global magma ocean because the melt would crystallize before it could spread over the surface, as discussed below. The timescale of radial relaxation decreases significantly as the volume of the melt pool become large (as in the case of a giant impact), especially in the case of a non-Newtonian rheology for the solid part of the proto-planet, which makes the rapid formation of a global magma ocean by isostatic re-adjustment following a giant impact highly likely.

Cooling times of magma oceans

Some of the impacts as calculated from the N-body accretion models may occur on a molten surface if a global magma ocean persists for a longer period of time. To determine whether this is the case, a simple model is used to estimate the cooling time of a global magma ocean. This model is described by the following equation:

$$4\pi R^2 F = \frac{dr}{dt}\rho L_m 4\pi r^2 - \frac{4}{3}\pi\rho C_p \frac{d}{dt}\left[T\left(R^3 - r^3\right)\right]$$ (4)

which balances the heat flux, $F$, through the planet's surface on the left hand side, with the latent heat, $L_m$, that is released during crystallisation (first term on the right) and the heat released through secular cooling (second term on the right) (Elkins-Tanton, 2008). For symbol explanation, see Table 1. Here the heat flux is approximated by a constant value, which is certainly a simplification. However, what really matters is whether the solidification timescale is large or small compared to the time interval between giant impacts. The latter parameter is provided by the accretion simulations.

If cooling occurs via radiation from a deep magma ocean, without an insulating atmosphere or proto-crust, the Rayleigh number is extremely high ($10^{27}$-$10^{32}$), partly because of the extremely low viscosity of peridotite melt (Liebske et al., 2005). Thus, convection is in the hard turbulent regime with convective velocities of several meters per second. The heat flux is extremely high and the solidification timescale, at least for the deeper (lower-mantle) part of the magma ocean, is on the order of $10^3$ years (Solomatov, 2000, 2015; Rubie et al., 2003) – much shorter than the time intervals between impacts. On the other hand, if a solid crust forms, the cooling rate is limited by conduction through this lid and mantle solidification could then take on the order of $10^8$ years – much longer than the intervals between impacts. An intermediate case arises when a thick steam atmosphere is present, in which case the heat flux is limited to 300-500 W/m$^2$ and is set by the thermodynamic properties of the atmosphere (Sleep et al. 2014). In this case the solidification timescale can be comparable to the impact interval and here the assumption of a constant heat flux is quite good, because it depends on atmospheric and not surface or interior properties. Below we present the results of calculations assuming either a thick steam atmosphere (F=475 W/m$^2$) or a radiating magma ocean (F=2×10$^5$ W/m$^2$). The former value was determined by fitting the cooling times of Lebrun et al. (2013) with our cooling model and is of the same order of magnitude as values determined for the maximum heat flux through a dense atmosphere (Sleep et al., 2014; Hamano et al., 2013). The latter value was determined using the same fitting procedure for a body without an atmosphere; the result is similar to the radiation of a black body with a surface temperature of ~1000 °C. Note that the constant heat fluxes used in our models represent time-averaged heat flows. In reality, heat fluxes will initially be relatively high and will decrease with time.

## Results and discussion

### Melt production

For each impact in an N-body accretion simulation, the impact angle, impact velocity and the masses of the two bodies involved are used to calculate melt production using the equations listed above. The impact histories from three different N-body simulations are used here. These are 4:1-0.25-7, 4:1-0.5-8, and i4:1-0.8-4 (Jacobson & Morbidelli, 2014; Rubie et al., 2015b) and were chosen because they result in realistic Earth-mass planets at ~1 AU (see Fig. 1 in Rubie et al., 2015). Here "4:1" designates the ratio of the total mass of all the embryos to the total mass of all the planetesimals at the beginning of the simulation. The middle term (e.g. "0.25") is the initial mass of embryos as a fraction of one Mars mass. For each set of starting parameters, as defined by these two terms, 10 simulations were run with very slight variations in the initial orbital characteristics of the starting bodies. The last term is the run number within the set of 10 simulations. The "i" in the third simulation number indicates that the starting mass of embryos increases with increasing heliocentric distant (from 0.7 to 3.0 AU) – see Table 3. The main difference between the three simulations used here is the initial mass of the embryos which ranges from 0.2 to 0.8 times the mass of Mars. Table 3 provides an overview of the initial parameters of the N-body models. Each simulation produces several final terrestrial planets and we define "Earth-like" planets as those that form close to 1 AU and have a mass close to one Earth mass (Rubie et al., 2015b).

Once the depth of melting for each impact has been calculated, the pressure and temperature at that depth is determined. Our ultimate aim is to use such conditions, and those at the base of subsequent global magma oceans, to model each episode of core formation and the resulting siderophile element partitioning throughout Earth's accretion. In the future, such models can potentially distinguish which N-body models most closely resemble Earth's accretion history if the calculated melt pressures for different N-body models are significantly different.

Figure 2 shows the pressure at the maximum melting depth for each impact, normalised by the core-mantle boundary pressure at the time of impact, $P_{CMB}$, as a function of the fraction of mass accreted to the final Earth-like planet. Closed symbols indicate results for giant impacts. Most of these collisions result in a melt pool that extends into the core. In this case, metal-silicate equilibration would take place close to the core-mantle boundary. However, most embryo-embryo collisions that result in shallower melt pools still melt the deep mantle with equilibration taking place at pressures close to the CMB pressure.

Planetesimal impacts nominally create shallower melt pools than embryo impacts with maximum pressures that range from 0.4 to 1.2 times the core-mantle boundary pressure. A few high-velocity impacts create even deeper melt pools. However, the results shown in Fig. 2 for planetesimals are not realistic. This is because each "planetesimal" in the N-body simulations is actually a tracer that is used to represent a swarm of much smaller bodies in order to reduce calculation times (O'Brien et al., 2006). Actual planetesimals likely have diameters of the order of 100 km. Impacts of bodies of this size will create only small amounts of melting and this is taken into account in the following discussion and the results that are presented below.

There are two end-member scenarios for planetesimal impacts. (1) They impact a solid surface due to the rapid crystallisation of a preceding magma ocean. In this case there will be little melting and the cores of such bodies will only segregate to the Earth's core when the next giant impact occurs and produces extensive melting. (2) Small bodies impact a pre-existing global magma ocean, created by an earlier giant impact, and metal-silicate equilibration takes place near the base of this magma ocean. In this case, the unrealistically large size of planetesimals will not significantly change the results, because the effect of a large number of small bodies impacting in a magma ocean around the same time will be similar to the effect of one large body with the same total mass impacting the magma ocean.

## Magma ocean crystallisation

To determine if some planetesimal impacts occur onto a pre-existing global magma ocean, we use the simple magma ocean crystallisation model described above (Eq. 4). If the time until the next impact on the proto-Earth is shorter than the crystallisation time of the magma ocean, the next impact will occur on a molten surface. In general, at the start of accretion the time intervals between impacts are relatively short. Most of these impacts are small and will cause only small amounts of melting. However, some of these collisions are between two similar-sized bodies resulting in a deep magma ocean that may not crystallise before the next impact.

The cooling model for magma oceans described above (Eq. 4), based on a constant surface heat flux, is used to estimate the crystallization time of a global magma ocean. To calculate the magma ocean depth from the volume of the spherical impact-induced melt pool, the silicate melt volume is spread out geometrically over the entire surface of the new body after collision and accretion. Since the cooling of a magma ocean depends strongly on the presence or absence of an insulating atmosphere, the results of the two end-member cooling models described above are presented here.

Magma ocean crystallization times are plotted as a function of the time until the next impact in Figure

3. Only embryo-embryo collisions that form the final Earth-like planet are shown. In the absence of an insulating atmosphere, almost all magma oceans crystallise before the next impact (almost all points plot above the dotted 1:1 line). When a dense insulating atmosphere is present, the majority of magma oceans would still be present when the next impact occurs.

## Metal-silicate equilibration depths

The magma ocean crystallisation model described above is used to determine the depth of the crystallizing magma ocean at the time of the next planetesimal impact. Based on the hydrodynamic model of Deguen et al. (2011), the metal from the impacting planetesimal can then be assumed to equilibrate with silicate liquid at this depth (Rubie et al. 2015b). In the case of an embryo impact onto a pre-existing magma ocean, the volume of the calculated melt pool that extends below the base of the magma ocean into previously solid mantle is added to the volume of the magma ocean to calculate a new magma ocean depth. Metal-silicate equilibration of the embryo material will likely occur at the bottom of the melt pool prior to isostatic readjustment.

Assuming that metal-silicate equilibration and segregation of metal to the proto-core occur only when a large volume of melt is created by an embryo-embryo collision (giant impact), we have the following possibilities for equilibration:

- *Embryo-embryo collisions*: Equilibration takes place at the bottom of the melt pool (or at the core-mantle boundary if the melt pool extends into the core) before a global magma ocean is created by isostatic readjustment and lateral spreading. In this case the equilibration pressure will generally be close to the core-mantle boundary pressure (Figure 4, large symbols).

- *Planetesimal impact in a magma ocean that survives from a previous embryo-embryo collision*: Equilibration will take place at the base of the magma ocean. The depth of the magma ocean is dependent on the time between the embryo-embryo collision and the planetesimal impact as well as the cooling rate of the magma ocean (Figure 4, points contained in the black ellipse).

- *Planetesimal impact on a solid surface before the last giant impact*: In this case, there is insufficient melt at the time of impact for significant equilibration and metal-silicate equilibration. However, the next giant impact will mix this material with the target material. Therefore, equilibration takes place at the bottom of the deep melt pool created by the next giant impact (Figure 4, points contained in the magenta ellipse).

- *Planetesimal impact on a solid surface after the last giant impact (or in the case of no giant impacts)*: in this case there is no subsequent significant melting event available to equilibrate the material and little mixing of the impactor and target material will occur, the impactor material forms a late veneer (Figure 4, points contained in the red ellipse).

The evolution of metal-silicate equilibration pressures for the Earth-like planets in the three N-body accretion simulations and, in each case, for the two end-member cooling/atmosphere models are shown in Figure 5. Embryo-embryo collisions (large symbols) are visible as maxima, whereas the planetesimal impacts in a residual magma ocean result in a decreasing equilibration pressure with increasing time. Planetesimal impacts onto a solid surface result in a plateau at a pressure corresponding to the depth of the next magma ocean or a zero pressure when the material is late veneer and there is no subsequent giant impact.

As expected, the fast-cooling models without an insulating atmosphere mainly show plateaus because magma oceans crystallize before the next planetesimals impact the surface of the planet. The equilibration pressures for almost all collisions are then dominated by the initial depths of the melt pools that are created by embryo-embryo collisions (i.e. giant impacts). These pressures are mostly higher than the equilibration pressures in the slow-cooling models, where many planetesimals equilibrate in partially crystallised magma oceans.

The differences in equilibration pressures between the slow-cooling and fast-cooling models are greatest when the time between embryo-embryo collisions is just long enough for the magma ocean to crystallise in the slow-cooling models. This results in many small impacts in a partially crystallised magma ocean as can be seen in Figure 5b. In models with smaller embryo masses (model 4:1-0.25-7, Figure 5a) there are more embryos and therefore more embryo-embryo collisions. This results in embryo impacts into partially-crystallised magma oceans, resulting in magma oceans that reach almost to the core-mantle-boundary. Equilibration pressures become similar to equilibration pressures in the fast cooling model where most of the equilibration takes place at or near the core-mantle boundary.

The dotted lines in Figure 5 indicate the equilibration pressures determined by Rubie et al. (2015b) of ~$0.7P_{CMB}$ (where $P_{CMB}$ is the CMB pressure at the time of each impact). Our calculations show that equilibration at the bottom of melt pools created by giant impacts likely occur at higher pressures. This means that if there was no atmosphere around a growing planet equilibration pressures will generally be >$0.7P_{CMB}$. However, in the case of a planet with a dense insulating atmosphere, many of the planetesimal impacts in a slow-cooling magma ocean equilibrate at lower pressures. This scenario is therefore most consistent with the results of Rubie et al. (2015b).

Discussion

The planetesimals in the N-body accretion models used here are unrealistically large and are actually tracers that represent swarms of much smaller bodies. However, since small planetesimal impacts do not create enough melt to cause a significant equilibration event, these bodies equilibrate either in a pre-existing magma ocean or at the bottom of the melt pool created by a subsequent giant impact. The exact size and timing of the planetesimal impactors therefore have little or no influence on the equilibration pressure.

Because of a number of simplifying assumptions, the results presented here must be considered to be preliminary. For example, the simple analytical model of melt production may not be accurate (Marchi et al., 2014) and does not take into account impacts onto existing magma oceans. In addition, the cooling models, based on constant time-averaged heat fluxes, are extremely simple. At this stage, our main intention is to emphasize the potential of this novel approach for modelling multistage core formation.

## Conclusions

Our results show that metal-silicate equilibration pressures during core formation are likely to depend strongly on the rate of magma ocean crystallisation compared to the time interval between impacts. On a planet with a dense atmosphere, or one which develops a solid crust, magma oceans will cool slowly and numerous small impactors will equilibrate at the bottom of this magma ocean. If the liquid magma radiates heat directly to space, most magma oceans will have crystallised before the next impact and then planetesimal material generally only equilibrates when the next giant impact creates a large and deep melt volume.

If Earth lacked an insulating atmosphere and/or a solid crust during accretion, metal-silicate equilibration would have mostly taken place at core-mantle boundary pressures. This is unlikely because, based on the partitioning of siderophile elements between the core and mantle, average equilibration pressures must have been considerably lower (e.g. Wood et al., 2006; Righter, 2011; Rubie et al., 2015a, 2015b). Thus, based on the present results, the Earth most likely had magma oceans with lifetimes comparable to or longer than the interval between impacts. A single, long-lived magma ocean is a possibility, especially if a solid crust developed. This possibility, however, is not favoured based on noble gas isotopic arguments which have been used to argue for several generations of magma oceans (Tucker and Mukhopadhyay, 2014). Conversely these isotopic results are consistent with an Earth possessing a thick, insulating atmosphere for much of its accretion history.

The accretion history also strongly influences the equilibration pressures for slow-cooling models. The three N-body simulations used for this study started with different sized and different numbers of embryos. When there are fewer embryo-embryo collisions, there are more planetesimal impacts in global magma oceans. This results in lower equilibration pressures throughout most of the planet's growth history. This difference may help determine which scenarios are most realistic for the Earth by combining N-body simulations with core formation models in which equilibration pressures are calculated as in this study.

**Acknowledgements.** J.deV, S.A.J., A.M. and D.C.R. were supported by the European Research Council Advanced Grant "ACCRETE" (contract number 290568). We thank Sarah Stewart for discussions and two anonymous reviewers and the associate editor (B. Mysen) for their constructive comments.

## References

Abramov O, Wong SM, Kring DA (2012) Differential melt scaling for oblique impacts on terrestrial planets. Icarus 218:906-916

Aitta A (2006) Iron melting curve with a tricritical point. J Stat Mech P12015

Barr AC, Citron RI (2011) Scaling of melt production in hypervelocity impacts from high-resolution numerical simulations. Icarus 211:913-916.

Benz W, Cameron AGW, Melosh HJ (1989) The origin of the Moon and the single impact hypothesis. III. Icarus 81:113-131.

Bjorkman MD, Holsapple KA (1987) Velocity scaling impact melt volume. Int J Impact Eng 5:155-163

Canup RM (2012) Forming a Moon with an Earth-like composition via a giant impact. Science 338:1052-1055

Canup RM, Asphaug E (2001) Origin of the Moon in a giant impact near the end of the Earth's formation, Nature 412:708-712

Chambers JE (2001) Making more terrestrial planets. Icarus 152:205-224

Chambers JE (2013) Late-stage planetary accretion including hit-and-run collisions and fragmentation. Icarus 224:43-56

Chambers JE, Wetherill GW (1998) Making the terrestrial planets: N-body integrations of planetary embryos in three dimensions. Icarus 136:304-327

Clauser C (2011) Thermal storage and transport properties of rocks, I: Heat capacity and latent heat. In: Gupta H (ed) Encyclopedia of Solid Earth Geophysics. Springer, Dordrecht

Corgne A, Siebert J, Badro J (2009) Oxygen as a light element: A solution to single-stage core formation. Earth Planet. Sci. Lett. 288:108–114

Ćuk M, Stewart ST (2012) Making the Moon from a fast-spinning Earth: A giant impact followed by resonant despinning. Science 338:1047-1052.

Dahl TW, Stevenson DJ (2010) Turbulent mixing of metal and silicate during planet accretion – An interpretation of the Hf-W chronometer. Earth Planet. Sci. Lett. 295, 177-186.

Deguen R, Olson P,  Cardin P (2011) Experiments on turbulent metal- silicate mixing in a magma ocean, Earth Planet. Sci. Lett., 310:303-313.

Deguen R, Landeau M, Olson P (2014) Turbulent metal-silicate mixing, fragmentation, and equilibration in magma oceans. Earth Planet. Sci. Lett. 391, 274-287.

Elbeshausen D, Wünnemann, K., Collins, G. S. (2009) Scaling of oblique impacts in frictional targets: Implications for crater size and formation mechanisms. Icarus 204, 716-731.

Elkins-Tanton L (2008) Linked magma ocean solidification and atmospheric growth for Earth and Mars. Earth Planet Sci Lett 271:181-191

Greenberg R, Hartmann WK, Chapman CR, Wacker, JF (1978) Planetesimals to planets—Numerical simulation of collisional evolution. Icarus 35:1-26.

Hartmann, W.K. and Davis, D. R. (1975) Satellite-sized planetesimals and lunar origin. Icarus 24, 504-515.

Hamano K, Abe Y, Genda H. (2013) Emergence of two types of terrestrial planet on solidification of magma ocean. Nature 497:607-610.

Jacobson, S. A., Morbidelli, A. (2014) Lunar and terrestrial planet formation in the Grand Tack scenario. Philosophical Transactions of the Royal Society A 372:2024, 20130174.

Jacobson SA, Morbidelli A, Raymond SN, O'Brien DP, Walsh KJ, Rubie DC (2014) Highly siderophile elements in the Earth's mantle as a clock for the Moon-forming impact. Nature 508:84-87

Jacobson, S. A. and Walsh, K. J. (2015) Earth and Terrestrial Planet Formation, in The Early Earth: Accretion and Differentiation (eds J. Badro and M. Walter), John Wiley & Sons, Inc, Hoboken, NJ.

Johansen, A., Jacquet, E., Cuzzi, J. N., Morbidelli, A., Gounelle, M.. (2015). New Paradigms For Asteroid Formation. in Asteroids IV (ed. Michel, P., DeMeo, F., and W. Bottke) University of Arizona Press Space Science Series

Kendall JD, Melosh HJ (2016) Differentiated planetesimal impacts into a terrestrial magma ocean: Fate of the iron core. Earth Planet. Sci. Lett. submitted.

Kokubo E, Ida S (1998) Oligarchic growth of protoplanets. Icarus 131:171-178

Lambrechts M, Johansen A (2012) Rapid growth of gas-giant cores by pebble accretion. Astronomy and Astrophysics 544:A32.

Lebrun T, Massol H, Chassefière E, Davaille A, Marcq E, Sarda P, Leblanc F, Brandeis G (2013) Thermal evolution of an early magma ocean in interaction with the atmosphere. J Geophys Res Planets 118:1155-1176

Li J, Agee CB (1996) Geochemistry of mantle-core formation at high pressure, Nature 381:686-689.

Lide DR (ed) (1995) CRC handbook of chemistry and physics. CRC Press, USA

Liebske C, Schmickler B, Terasai H, Poe BT, Suzuki A, Funakoshi K, Ando R, Rubie DC (2005) Viscosity of peridotite liquid up to 13 GPa: Implications for magma ocean viscosities. Earth Planet. Sci. Lett. 240:589-604.

Liebske C, Frost DJ (2012) Melting phase relations in the MgO-MgSiO3 system between 16 and 26 GPa: Implications for melting in Earth's deep interior. Earth Planet Sci Lett 345-348:159-170

Kleine T, Touboul M, Bourdon B, Nimmo F, Mezger K, Palme H, Stein SB, Yin Q-Z, Halliday AN (2009) Hf-W chronology oft he accretion and early evolution of asteroids and terrestrial planets. Geochim. Cosmochim. Acta 73:5150-5188.

Marchi S, Bottke WF, Elkins-Tanton LT, Bierhaus M, Wuennemann K, Morbidelli A, Kring DA (2014) Widespread mixing and burial of Earth's Hadean crust by asteroid impacts. Nature 511:578-582.

Marinova MM, Aharonson O, Asphaug E (2011) Geophysical consequences of planetary-scale impacts into a Mars-like planet. Icarus 211:960-985.

McDonough WF (2003) Compositional model for the Earth's core, in: Carlson RW (Ed), Treatise on Geochemistry, Volume 2-The Mantle and Core. Elsevier-Pergamon, Oxford, pp. 547-568.

Monnereau M, Yuen DA (2002) How flat is the lower-mantle temperature gradient? Earth Planet. Sci. Lett.202, 171-183.

Navrotsky A (1995) Thermodynamic properties of minerals. In: Ahrens TJ (ed) Mineral physics and crystallography. American Geophysical Union, pp. 18-28.

Nakajima M, Stevenson DJ (2015) <u>Melting and mixing states of the Earth's mantle after the Moon-forming impact</u>. Earth Planet. Sci. Lett. 427:286-295.

Nimmo F, O'Brien DP, Kleine T (2010) Tungsten isotopic evolution during late-stage accretion: Constraints on Earth-Moon equilibration. Earth Planet. Sci. Lett. 292:363-370.

O'Brien DP, Morbidelli A, Levison HF (2006) Terrestrial planet formation with strong dynamical friction. Icarus 184:39-58

O'Brien, D. P., Walsh, K. J., Morbidelli, A., Raymond, S. N., Mandell, A. M. (2014) Water delivery and giant impacts in the Grand Tack scenario. Icarus 239:74-84

Palme, H., O'Neill, H.St.C., 2013. Cosmochemical estimates of mantle composition. In: Carlson, R.W. (Ed.), Treatise on Geochemistry – The Mantle and Core, second ed., vol. 2. Elsevier-Pergamon, Oxford, pp. 1–39.

Pierazzo E, Melosh HJ (2000) Melt production in oblique impacts. Icarus 145:252-261

Pierazzo E, Vickery AM, Melosh HJ (1997) A reevaluation of impact melt production. Icarus 127:408-423

Reese CC, Solomatov VS (2006) Fluid dynamics of local Martian magma oceans. Icarus 184:102-120

Righter K (2011) Prediction of metal–silicate partition coefficients for siderophile elements: An update and assessment of PT conditions for metal–silicate equilibrium during accretion of the Earth. *Earth*

*Planet. Sci. Lett.* 304:158-167.

Rubie DC, Melosh HJ, Reid JE, Liebske C & Righter K (2003) Mechanisms of metal-silicate equilibration in the terrestrial magma ocean. Earth Planet. Sci. Lett. 205:239-255.

Rubie DC, Frost DJ, Mann U, Asahara Y, Nimmo F, Tsuno K, Kegler P, Holzheid A, Palme H (2011) Heterogeneous accretion, composition and core-mantle differentiation of the Earth. Earth Planet Sci Lett 301:31-42

Rubie DC, Nimmo F, Melosh HJ (2015a) Formation of Earth's core. *In*: Treatise on Geophysics Vol. 9: Evolution of the Earth, DJ Stevenson (ed.), 2$^{nd}$ Edition, edited by G Schubert, Elsevier, Amsterdam, pp. 43-79.

Rubie DC, Jacobson SA, Morbidelli A, O'Brien DP, Young ED, de Vries J, Nimmo F, Palme H, Frost DJ (2015b) Accretion and differentiation of the terrestrial planets with implications for the compositions of early-formed Solar System bodies and accretion of water. Icarus 248:89-108

Rudge JF, Kleine T, Bourdon B (2010) Broad bounds on Earth's accretion and core formation constrained by geochemical models. Nature Geoscience 3:439-443.

Samuel H (2012) A re-evaluation of metal diapir breakup and equilibration in terrestrial magma oceans. Earth Planet. Sci. Lett. 313-314:105-114.

Schmidt, R. M., Housen, K. R. (1987) Some recent advances in the scaling of impact and explosion cratering. Int. J. Impact Eng. 5:543-560.

Sleep NH, Zahnle KJ, Lupu, RE (2014) Terrestrial aftermath of the Moon-forming impact. Phil. Trans. R. Soc. A 372:20130172

Solomatov V S, 2000, Fluid dynamics of a terrestrial magma ocean, in *Origin of the Earth and Moon*, Canup R M and Righter K, eds., pp. 323-338, Univ. Ariz. Press.

Solomatov VS (2015) Magma oceans and primordial mantle differentiation. *In*: Treatise on Geophysics Vol. 9: Evolution of the Earth, DJ Stevenson (ed.), 2nd Edition edited by G Schubert, Elsevier, Amsterdam, pp. 81-104.

Stevenson DJ (1990) Fluid dynamics of core formation, in the *Origin of the Earth*, Newsom HE and Jones JH, eds., Oxford Univ. Press.

Stewart ST, Lock SJ, Mukhopadhyay S (2014) Atmospheric loss and volatile fractionation during giant impacts. Lunar and Planetary Science Conference 45, Abstract 2869

Tonks, W.B., Melosh, H.J. (1992) Core formation by giant impacts. Icarus 100:326-346.

Tucker JM, Mukhopadhyay S (2014) Evidence for multiple magma ocean outgassing and atmospheric loss episodes from mantle noble gases. Earth Planet. Sci. Lett. 393, 254-265.

Wade J, Wood BJ (2005) Core formation and the oxidation state of the earth. Earth Planet. Sci. Lett. 236:78–95.

Walsh KJ, Morbidelli A, Raymond SN, O'Brien DP, Mandell AM (2011) A low mass for Mars from Jupiters early gas-driven migration. Nature 475:206-209

Walter MJ, Cottrell E (2013) Assessing uncertainty in geochemical models for core formation in Earth. Earth Planet Sci Lett 365: 165-176

Wood BJ, Walter MJ, Wade J (2006) Accretion of the Earth and segregation of its core. Nature 441: 825-833

Wood, BJ, Wade J, Kilburn MR (2008) Core formation and the oxidation state of the Earth: Additional constraints from Nb, V and Cr partitioning. Geochim. Cosmochim. Acta 72:1415-1426.

**Figure Captions**

**Figure 1**: Isostatic readjustment and lateral spreading of an initially-spherical impact-induced melt pool (a) leading to global magma ocean formation (b). Due to the lower density of the melt (red) compared to the surrounding solid mantle (blue), the solid mantle rises isostatically beneath the melt pool (white arrows) by solid-state deformation and the melt spreads over the surface of the planetary embryo (black arrows).The proto-core is shown in yellow.

**Figure 2**: The pressure at the melting depth as a function of the mass of the growing planet for all accretional collisions that create an Earth-like planet. Results are shown for the three different N-body accretion models used in this study (Table 3). Melt pressures for planetesimal impacts are unrealistically high, because the planetesimals in the N-body models actually represent a swarm of numerous smaller bodies. Each of these smaller impacts would result in much smaller amounts of melting. Note that pressure is normalised to the core-mantle boundary pressure ($P_{CMB}$) at the time of each impact.

**Figure 3**: Time until the next impact versus magma ocean crystallisation time for Earth-building giant impacts (embryo-embryo collisions). All points above the 1:1 line represent magma oceans that crystallise before the next impact. In the case where there is no atmosphere (surface heat flux=200,000 W/m$^2$) almost all magma oceans crystallise before the next impact. When the presence of a dense atmosphere is considered (surface heat flux=475 W/m$^2$), almost all global magma oceans are still present when the next impact occurs.

**Figure 4**: Equilibration pressures calculated for the Earth forming impacts. Large symbols are embryo-embryo collisions - these giant impacts create enough melting for a core-formation event to occur and material equilibrates at the bottom of the melt pool, created by the impact. Small symbols are planetesimal impacts into a leftover magma ocean (MO) that equilibrate at the bottom of this magma ocean (e.g. black ellipse). A horizontal line indicates planetesimal collisions that occur on a solid surface and equilibrate at the time of the next giant impact (e.g. magenta ellipse). Impacts in the red

ellipse are planetesimal impacts on a solid surface after the last giant impact has occurred.

**Figure 5**: Evolution of metal-silicate equilibration pressures during Earth's accretion calculated for the three N-body accretion simulations. In each case, results are shown for slow cooling/crystallizing magma oceans due to the presence of an insulating atmosphere (red symbols) and fast cooling/crystallizing magma oceans due to an absence of an insulating atmosphere (black symbols). Equilibration pressures determined by Rubie et al. (2015) *($P_{equil} \approx 0.7 \times P_{CMB}$* for all impacts, irrespective of size) are shown for comparison.

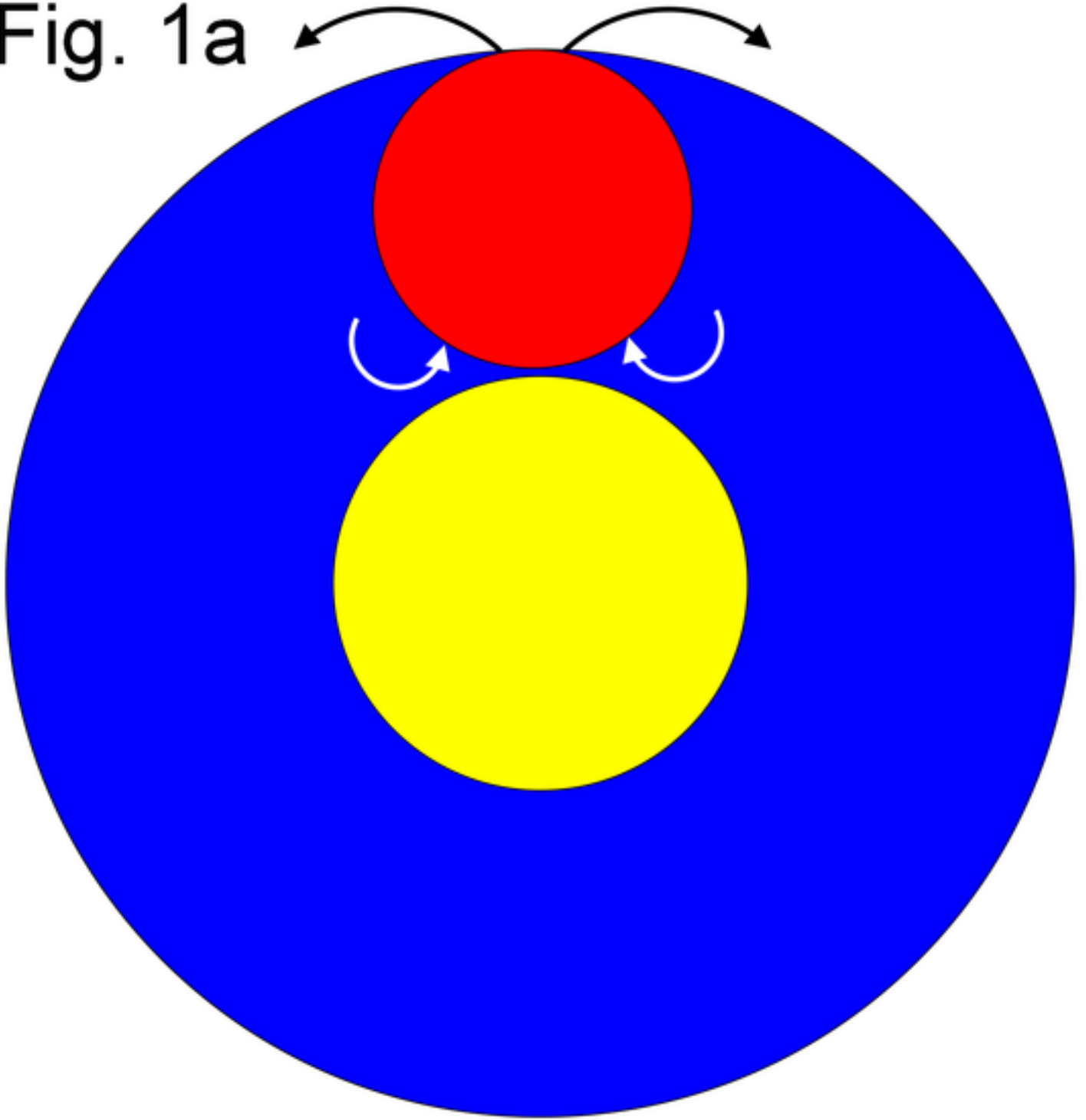

Fig. 1a

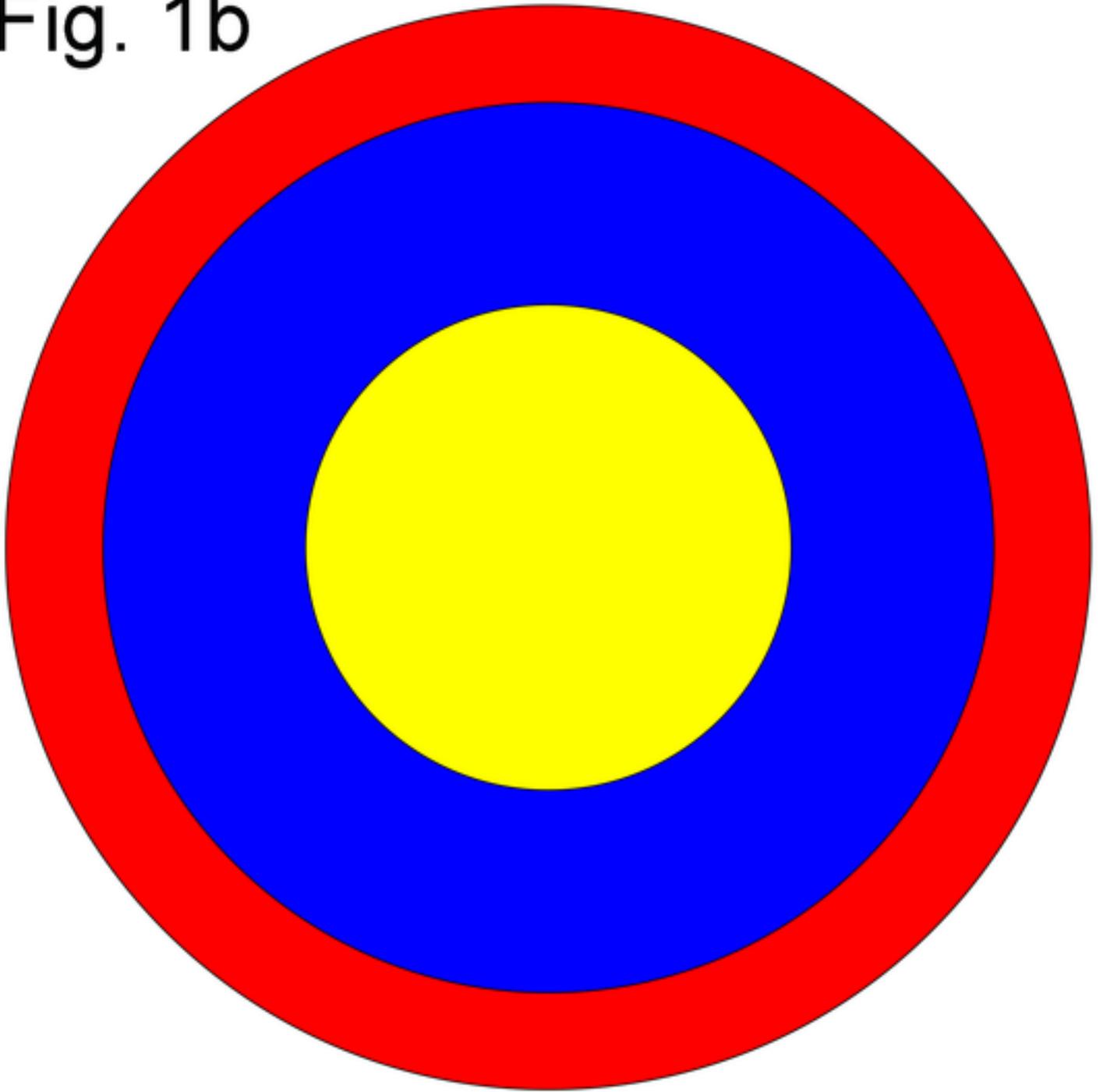

Fig. 1b

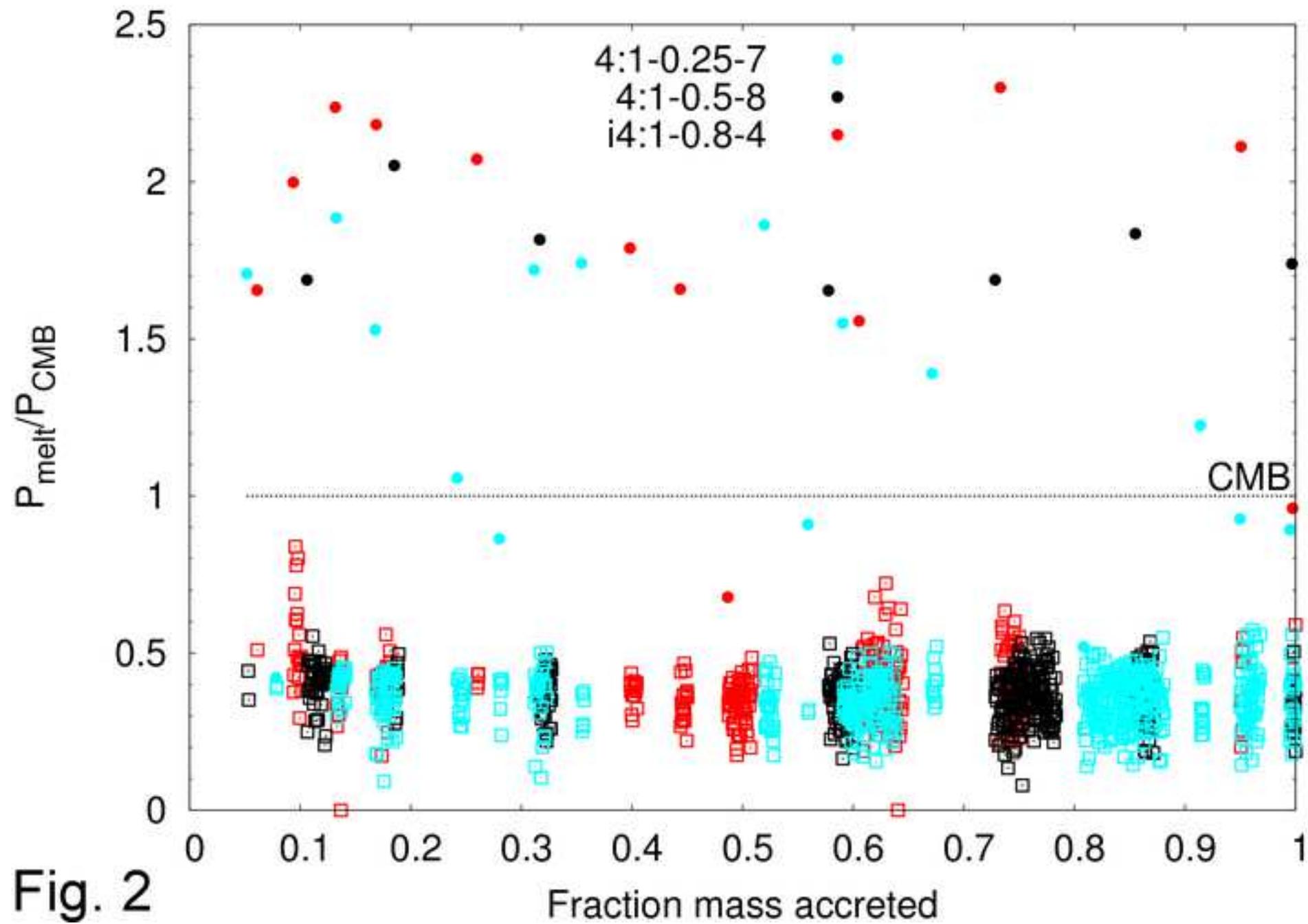

Fig. 2

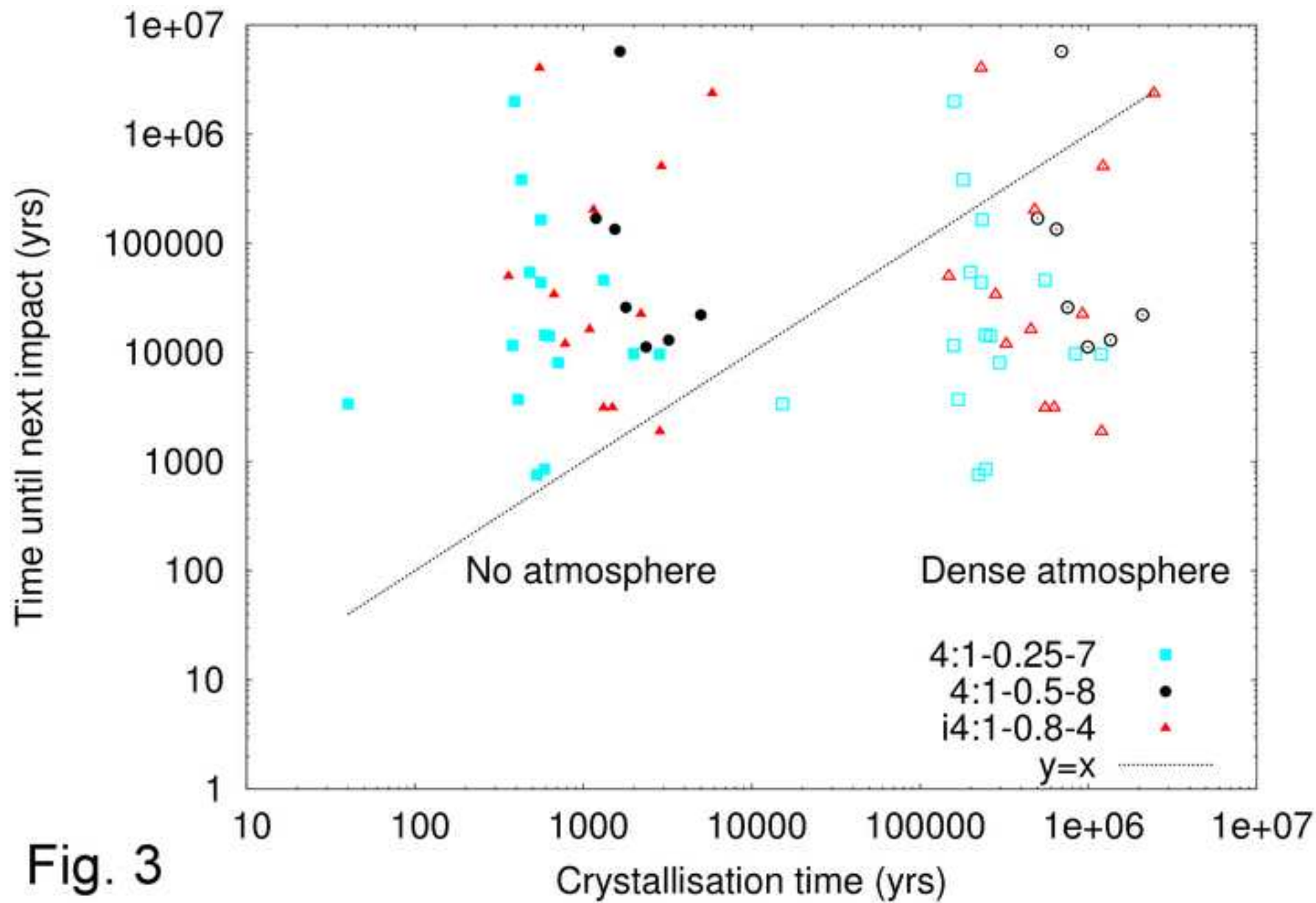

Fig. 3

Fig. 4

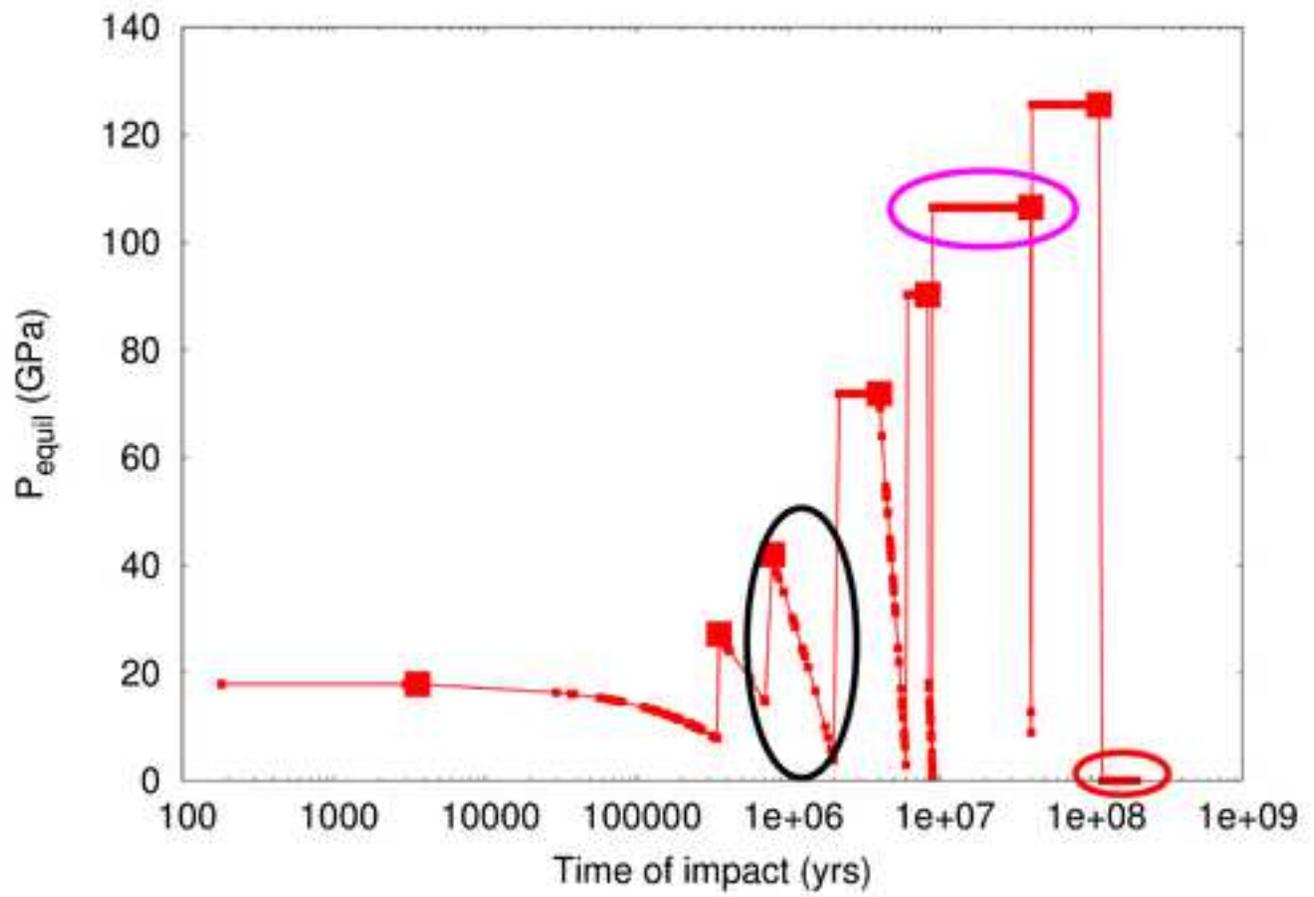

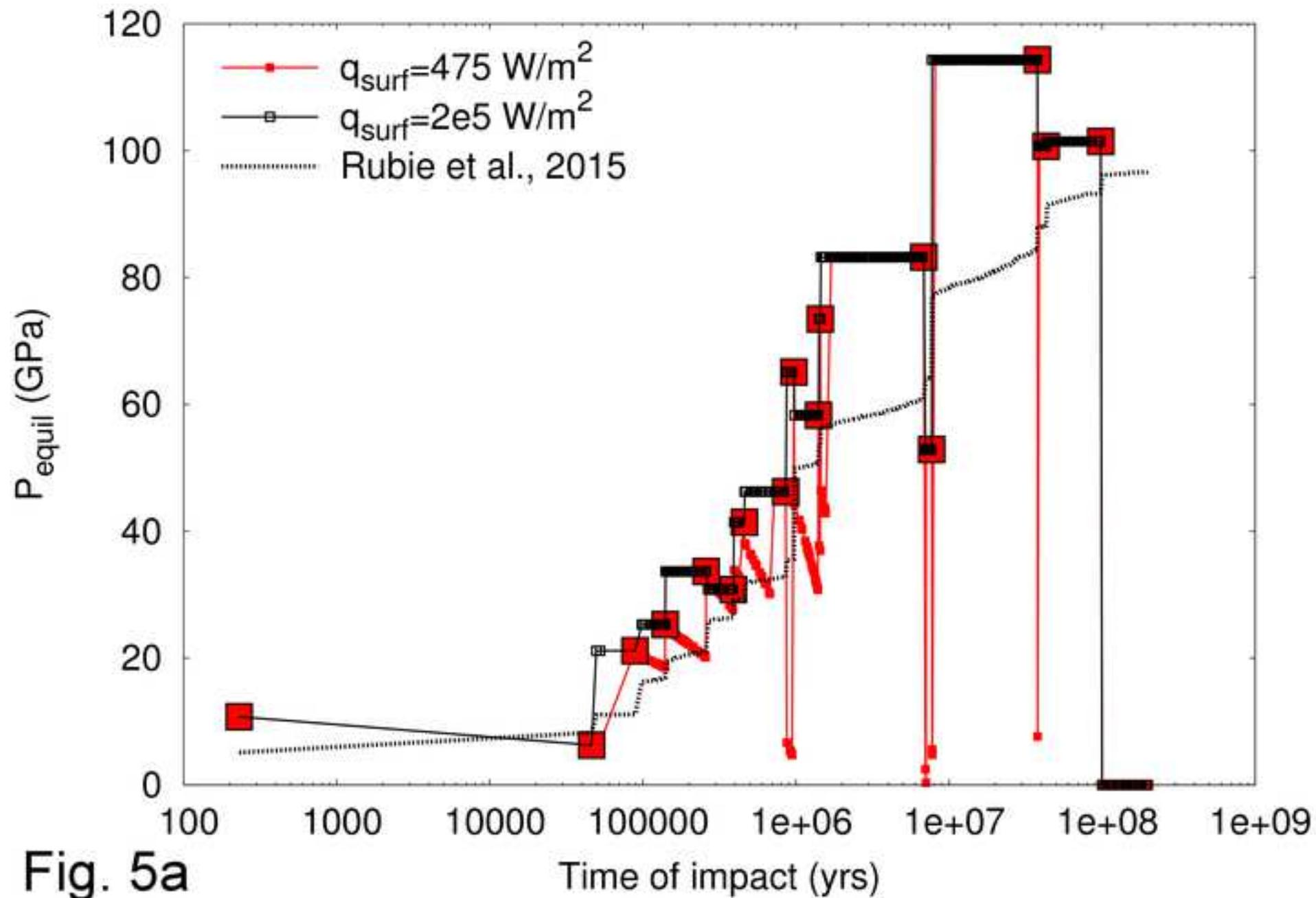

Fig. 5a

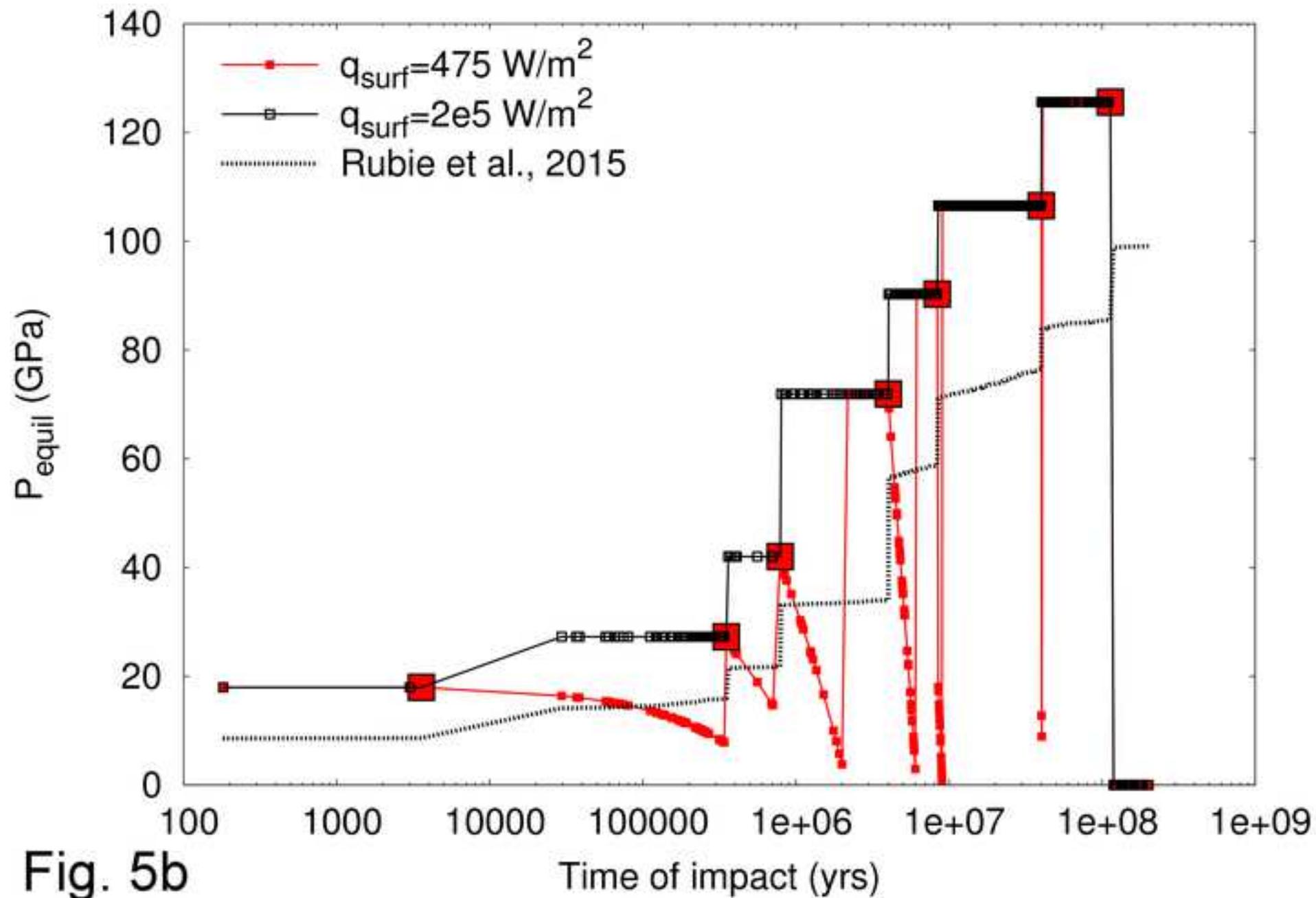

Fig. 5b

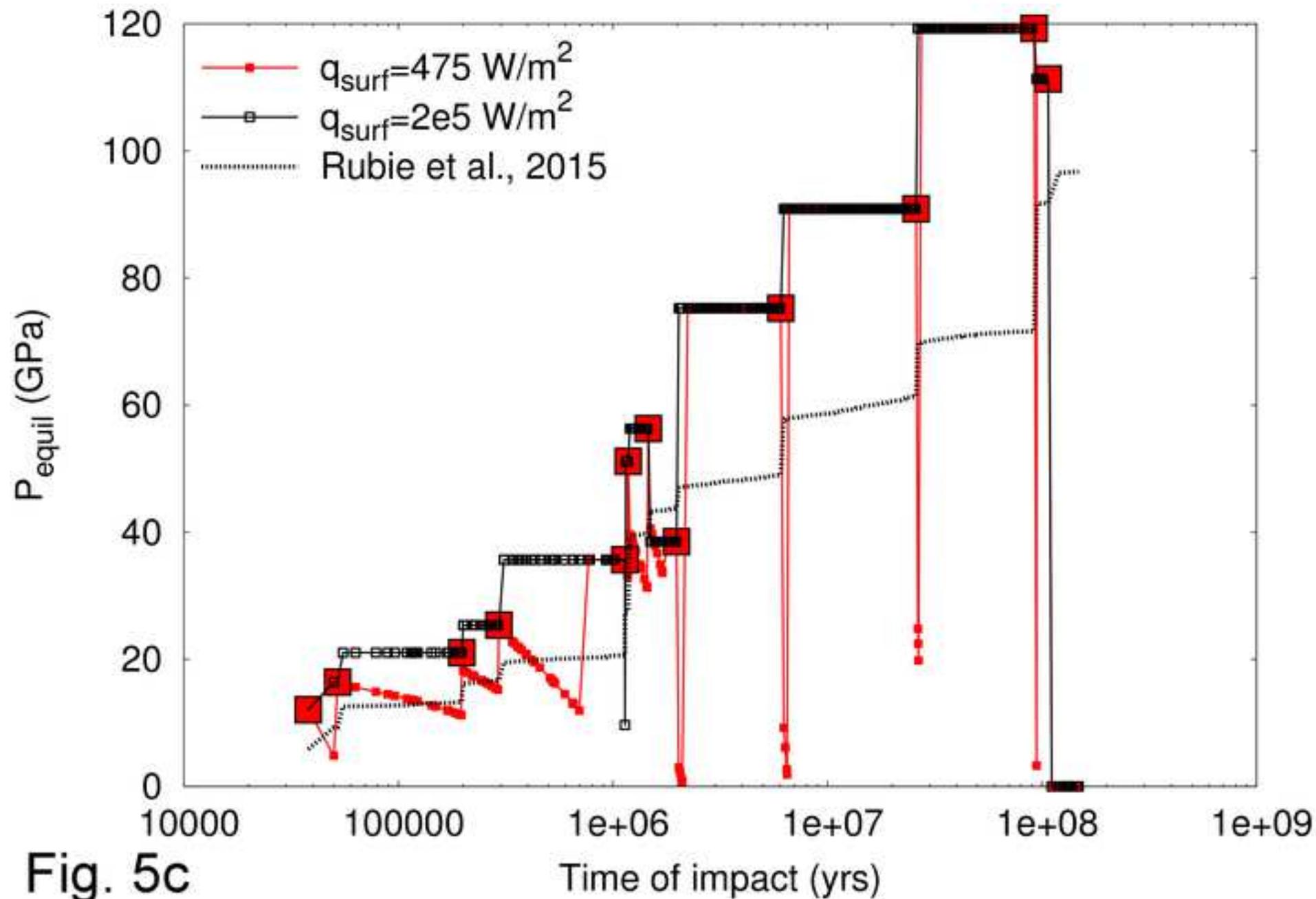

Fig. 5c

Table 1. Definition of the parameters used in the model description. Values that are constant throughout all models are listed. Values of the scaling constants $k$ and $\gamma$ are from Abramov et al. (2012).

| Symbol | Description | Value |
|---|---|---|
| $C_p$ | Specific heat | |
| $d_m$ | Depth of melting | |
| $D_p$ | Projectile diameter | |
| $E_m^0$ | Specific energy of melting | |
| $F$ | Surface heat flux | |
| $k$ | Experimentally determined scaling constant | 0.42 |
| $L_m$ | Latent heat of melting | |
| $M_{melt}$ | Mass of melt | |
| $m_p$ | Mass of projectile | |
| $T_{l0}$ | Liquidus temperature at the surface | |
| $T_l$ | Liquidus temperature | |
| $T_s$ | Surface temperature | 1750 K |
| $P$ | Pressure | |
| $R$ | Radius of body | |
| $r$ | Radial coordinate | |
| $v$ | Impact velocity | |
| $V_{melt}$ | Volume of melt | |
| $z$ | Depth coordinate | |
| $\gamma$ | π-group scaling parameter ($\gamma = 3\mu/(2+\mu)$) | 0.66 |
| $\theta$ | Impact angle | |
| $\mu$ | Energy/momentum scaling (velocity exponent) | 0.56 |
| $\rho_t$ | Target density | |
| $\rho_p$ | Projectile density | |

Table 2. Material properties

| Parameter | Dunite | Iron |
|---|---|---|
| $E_m^0$ | 9.0 MJ/kg (Pierazzo et al., 1997) | 10 MJ/kg (Pierazzo et al., 1997) |
| $C_p$ | 1300 J/kg/K (Clauser, 2011) | 449 J/kg/K (Lide, 1995) |
| $L_m$ | 718 kJ/kg (Navrotski, 1995) | 247 kJ/kg (Lide, 1995) |
| $T_{l0}$ | 1950 K (Liebske & Frost, 2012) | 1811 K (Aitta, 2006) |
| $dT_l/dP$ ($P$<60 GPa) | 28.3 K/GPa (Liebske & Frost, 2012) | 17.2 K/GPa (Aitta, 2006) |
| $dT_l/dP$ ($P$>60 GPa) | 14.0 K/GPa (Liebske & Frost, 2012) | 17.2 K/GPa (Aitta, 2006) |

Table 3. Starting parameters used for the three N-body accretion simulations (number and masses of embryos and planetesimals respectively). Collision histories of these three simulations are used for calculations of impact-induced melting in this study.

| Simulation: | 4:1-0.25-7 | 4:1-0.5-8 | i4:1-0.8-4 |
|---|---|---|---|
| No. of embryos | 170 | 87 | 82 |
| No. of planetesimals | 4346 | 4399 | 2760 |
| Embryo mass | $\sim 0.25 \times M_{Mars}$ | $\sim 0.5 \times M_{Mars}$ | $\sim 0.2\text{-}0.8 \times M_{Mars}$ [*] |
| Planetesimal mass | $< 0.04 \times M_{Mars}$ | $< 0.04 \times M_{Mars}$ | $< 0.04 \times M_{Mars}$ |

[*]Embryo mass increases with increasing heliocentric distance, from 0.7 to 3.0 AU.